\documentclass[11pt,twocolumn,twoside,a4paper,amsmath,amssymb,aps,showkeys,showpacs]{revtex4}

\usepackage{amsfonts}
\usepackage{graphics} 
\usepackage{epsfig}
\usepackage{fancyheadings}

\begin{document}
\thispagestyle{myheadings}
\rhead[]{}
\lhead[]{}
\chead[Cheuk-Yin Wong]{The Ridge Associated with the Near-side Jet}

\title{ 
The Ridge Associated with the  Near-side Jet
in High Energy-Heavy-Ion Collisions$^{\dagger}$}

\author{Cheuk-Yin Wong}

\affiliation{Physics Division, Oak Ridge National Laboratory, Oak
Ridge, TN\footnote{wongc@ornl.gov} 37831}

\affiliation{ Department of Physics and Astronomy, University of
Tennessee, Knoxville, TN 37996}


\begin{abstract}

The ridge particles associated with a near-side jet are identified as
medium partons kicked by the jet near the surface.  They carry direct
information on the parton momentum distribution at the moment of
jet-parton collisions and the magnitude of the longitudinal momentum
kick.  The extracted early parton momentum distribution has a rapidity
plateau structure with a thermal-like transverse momentum
distribution.  Such a rapidity plateau structure may arise from
particle production in flux tubes, as color charges and anti-color
charges separate at high energies.
\end{abstract}



\maketitle

\renewcommand{\thefootnote}{\fnsymbol{footnote}}

{\footnotetext[2]{Talk presented at the XXXIX International Symposium
on Multiparticle Dynamics, Gomel, Belarus, September 4-9, 2009.}}



\section{Introduction}
\label{introduction}

In central high-energy heavy-ion collisions, jets are produced in
nucleon-nucleon collisions and they interact with the dense matter
produced in the interacting region.  Depending on the degree of jet
attenuation and energy loss, jets of high $p_t$ particles can be
classified as near-side jets or away-side jets in the opposite
direction.

A near-side jet is characterized by the presence of associated
particles within a narrow cone along the trigger direction which
correspond to jet remnants of the high $p_t$ jet.  They retain the the
jet remnant characteristics as those in $pp$ and peripheral heavy-ion
collisions.  The near-side jet occurs when the high $p_t$ trigger
emerges near the surface of the produced parton medium.  In these
experiments, one measures the azimuthal angle $\phi$ and the
pseudorapidity angle $\eta$ of the jet and its associated particles,
from which one obtains the corresponding relative differences $\Delta
\phi=\phi_{\rm ass}-\phi_{\rm jet}$ and $\Delta \eta=\eta_{\rm ass}
-\eta_{\rm jet}$.  The probability distribution in $(\Delta \phi,
\Delta \eta)$ was found to be in the form of a ridge at
$\Delta\phi\sim 0$ running along the $\Delta \eta$ direction, in
addition to the narrow cone of the jet component at $ \Delta \phi\sim
0$ and $\Delta\eta \sim 0$.

Since the observation of the ridge associated with the near-side jet
by the STAR Collaboration \cite{Ada05,Put07,Wan07}, similar ridge
phenomena were observed by the PHOBOS \cite{Wen08} and PHENIX
Collaborations \cite{Ada08}.  Although many theoretical models have
been put forth to explain the ridge phenomenon
\cite{Won07,Won07a,Won08,Won08a,Won09a,Won09b,Shu07,Vol05,Chi08,Hwa03,Chi05,Pan07,Dum08,Gav08,Gav08a},
few provide direct comparisons with the extensive set of experimental
data.  Comparisons with experimental data over an extended region of
the associated particle phase space of $p_t$, $\Delta \phi$, and
$\Delta \eta$ have been carried out in the momentum kick model
\cite{Won07,Won07a,Won08,Won08a,Won09a,Won09b}.  By successfully
explaining the experimental data, the momentum kick model has
extracted a wealth of valuable information on the state of the dense
matter formed at the early stage of the collision process.

\section{The Momentum Kick Model}

In the momentum kick model, particles in the ridge component are
described as medium partons scattered (kicked) by the jet passing
through the medium.  Such a conclusion comes from the following
considerations.  It is observed that (i) the ridge yield increases
with the number of participants, (ii) the ridge yield is nearly
independent of the trigger jet properties, (iii) the baryon to meson
ratio of the ridge particles is more similar to those of the bulk
matter of the medium than those of the jet, and (iv) the slope
parameter of the transverse momentum distribution of ridge particles
is intermediate between those of the jet and the bulk matter
\cite{Ada05}-\cite{Lee08}. These features suggest that the ridge
particles are medium partons, at an earlier stage of the medium
evolution during the passage of the jet.  The azimuthal correlation of
the ridge particle with the jet suggests that the associated ridge
particle and the trigger jet are related by a collision.

As a result of the jet-parton collision, a medium parton received a
momentum kick ${\bf q}$ from the jet.  The momentum distribution of a
ridge parton at momentum ${\bf p}$ is then the momentum distribution
of the parton before the jet-parton collision at the shifted initial
momentum ${\bf p}_i={\bf p} - {\bf q}$.  The addition of the momentum
kick along the jet direction to the collided medium parton gives rise
to the peak in $\Delta \phi \sim 0$, and the extended pseudorapidity
distribution of the associated particles on the ridge arises from the
initial parton pseudorapidity distribution before the jet-parton
collision.  

The momentum distribution of associated particles in the `jet
component' at $\Delta \phi$=0 and $\Delta \eta$$\sim$0 in a
nucleus-nucleus collision can be assumed to be an attenuated
distribution of associated particles in a $pp$ collision, as expected
for production in an interacting medium near the surface.
Therefore, the total observed yield of associated particles per
trigger in A+A collisions consists of the sum of the jet and ridge
components,
\begin{eqnarray}
\label{obs}
& &\hspace*{-1.5cm}\left [ 
\frac{1}{N_{\rm trig}}
\frac{dN_{\rm ch}} 
{p_{t} dp_{t} d\Delta \eta  d\Delta \phi } \right ]_{\rm total}^{\rm AA} 
 \nonumber\\
& &
= 
\left [ f_J  \frac { dN_{\rm jet}^{pp}} {p_t dp_t\, d\Delta \eta\, d\Delta
\phi} \right ]_{\rm jet}^{\rm AA}
\nonumber \\
& &
+
\left [ f_R   \frac {2}{3} 
\langle N_k \rangle \frac { dF } {p_t dp_t\,
d\Delta \eta\, d\Delta \phi} \right ]_{\rm ridge}^{\rm AA} 
 \!\!\!\!,
\end{eqnarray}
where $ dN_{\rm jet}^{pp}/ p_t dp_t d\Delta \eta d\Delta \phi$ is the
distribution of associated particles in a $pp$ collision, $\langle N_k
\rangle$ is the average number of kicked partons per jet, and $dF/ p_t
dp_t d \eta\,d \phi $ is the normalized initial parton momentum
distribution before the jet-parton collisions, evaluated at the
shifted initial momentum ${\bf p}_i={\bf p}-{\bf
q}$.  Here $f_J$ and $f_R$ are the survival factors for the associated
particles, as they continue to pass through the medium after
production.

The experimental data calls for an early parton momentum distribution
that has a thermal-like transverse momentum distribution, modified for the
low-$p_t$ region, and a rapidity distribution that retains the
flatness at mid-rapidity but also respects the kinematic boundaries at
large rapidities and large $p_t$. Accordingly, we parametrize the
normalized initial parton momentum distribution as
\begin{eqnarray}
\label{dis2}
\hspace*{-0.3cm} 
\frac{dF}{ p_{ti}dp_{ti}dy_i d\phi_i}=
A_{\rm ridge} (1-x)^a 
\frac{ e^ { -\sqrt{m^2+p_{ti}^2}/T }} {\sqrt{m_d^2+p_{ti}^2}},
\end{eqnarray}
where $A_{\rm ridge}$ is a normalization constant, $x$ is the
light-cone variable
\begin{eqnarray}
\label{xxx}
x=\frac{\sqrt{m^2+p_{ti}^2}}{m_b}e^{|y_i|-y_b},
\end{eqnarray}
$a$ is the fall-off parameter that specifies the rate of decrease of
the rapidity distribution, $y_b$ is the beam parton rapidity, and
$m_b$ is the mass of the beam parton.  We expect $y_b$ to have a
distribution centered around the nucleon rapidity,
$y_N=\cosh^{-1}(\sqrt{s_{_{NN}}}/2m_N)$.  For lack of a definitive
determination, we shall set $y_b$ equal to $y_N$ and $m_b$ equal to
$m_\pi$.

\begin{figure} [h]
\includegraphics[angle=0,scale=0.40]{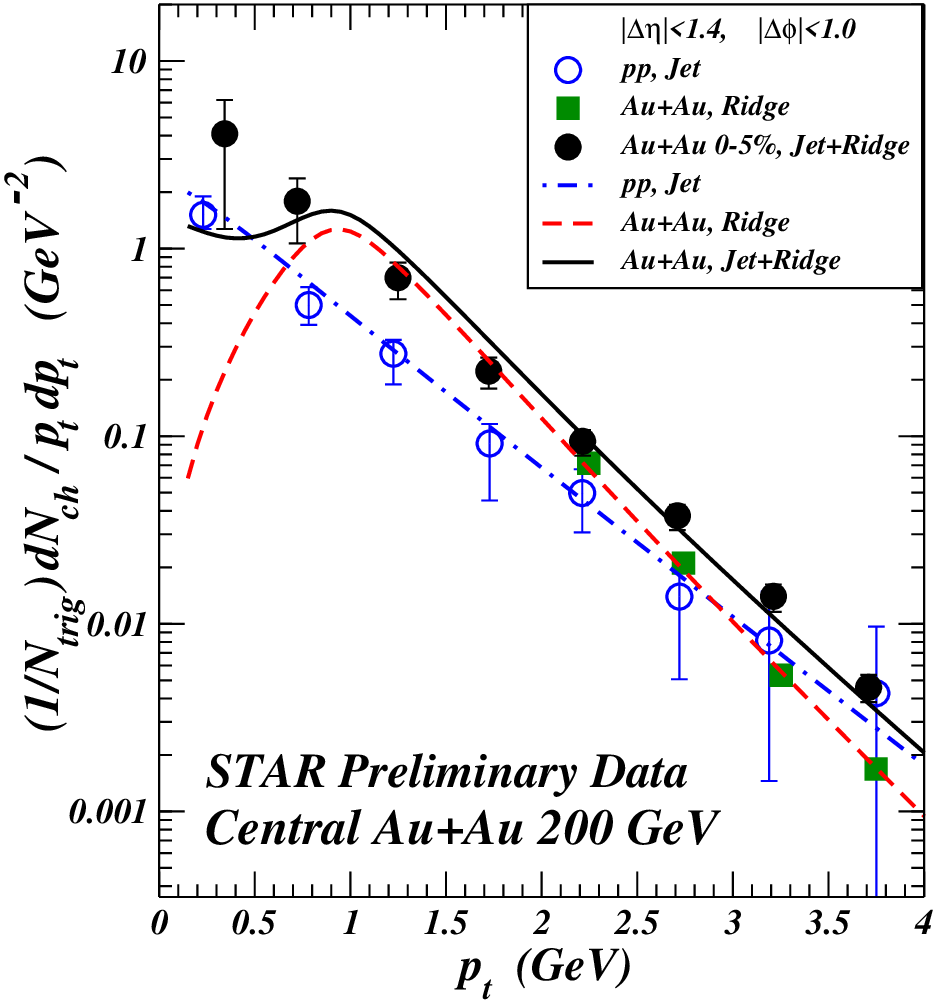}
\vspace*{0.0cm}
\caption{ The symbols represent STAR experimental data
\cite{Ada05,Put07} and the curves theoretical results of $dN_{\rm
ch}/N_{\rm trig}p_t dp_t$, for $pp$ and central AuAu collisions.}
\end{figure}

\section{Analysis of Experimental Associated Particle Data}

\begin{figure} [h]
\includegraphics[angle=0,scale=0.40]{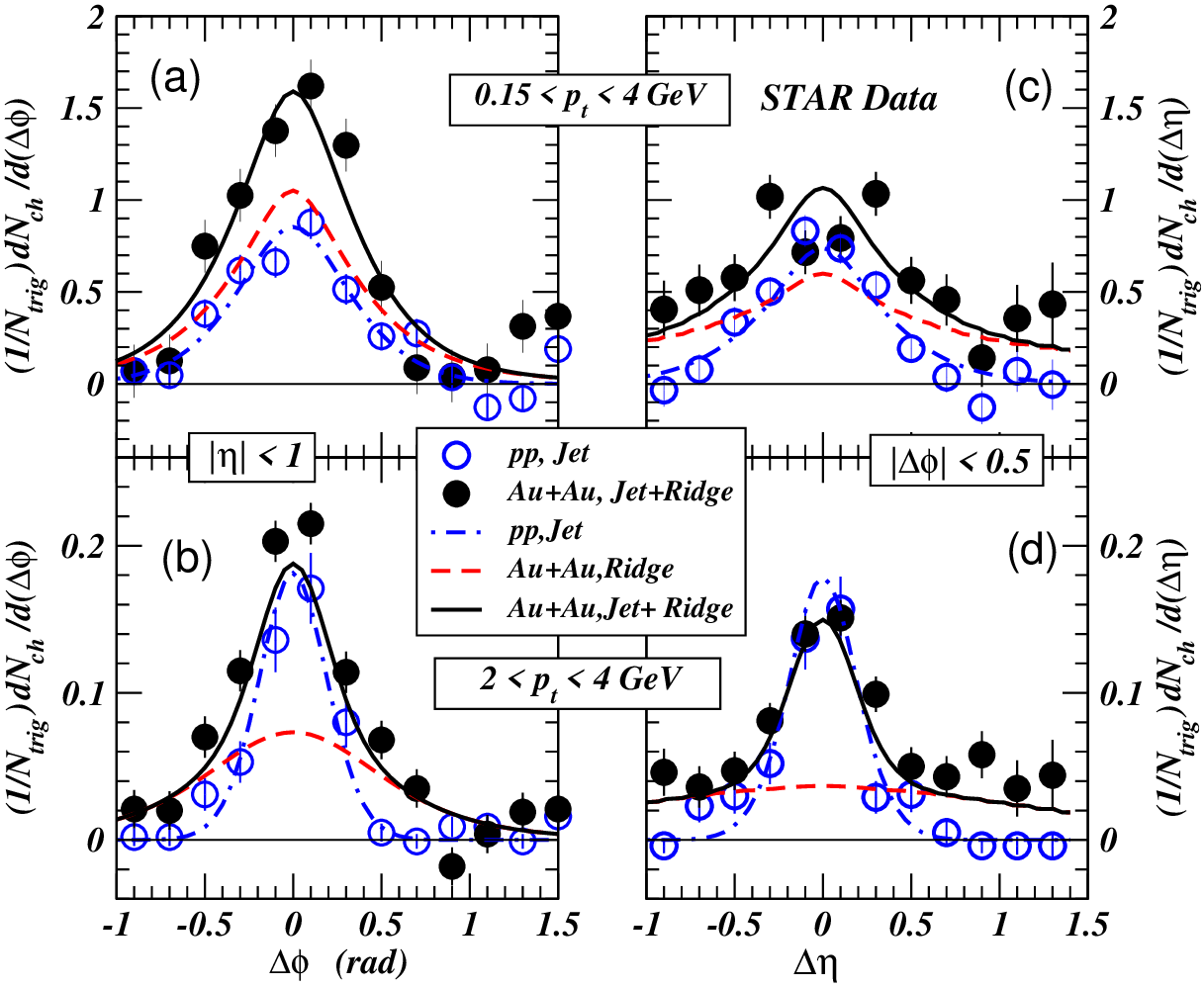}
\vspace*{0.0cm} 
\caption{ The symbols represent STAR experimental data \cite{Ada05}
and the curves theoretical results, for $pp$ and central AuAu
collisions.  (a) and (b) give the $dN_{\rm ch}/N_{\rm trig}d\Delta
\phi$ distributions.  (c) and (d) give the $dN_{\rm ch}/N_{\rm
trig}d\Delta \eta$ distributions.  }
\end{figure}

We analyze associated particles data for central Au-Au collisions and
$pp$ collisions at $\sqrt{s_{NN}}=200$ GeV.  The experimental data of
the near-side associated particles in $pp$ collisions, $ dN_{\rm
jet}^{pp}/ p_t dp_t d\Delta \eta d\Delta \phi$ in Eq.\ (\ref{obs}),
can be easily parametrized as a cone distribution with a cone width as
given in \cite{Won08a,Won09a}.  The near-side associated particle data
\cite{Ada05,Put07,Wen08,Ada08} for the experimental range of $p_t$,
$\Delta \phi$ and $\Delta \eta$ can be described by
\begin{eqnarray}
a=0.5, T=0.5 {\rm GeV},m_d=1.0{\rm GeV}, f_J=0.632.
\end{eqnarray}
The jet-medium interaction parameters depends on the centrality. They are 
\begin{eqnarray}
q_L= 1 {\rm ~GeV}, \, f_R \langle N_K \rangle = 3.8,
\end{eqnarray}
for STAR and PHOBOS data with 0-5\% and 0-10\% centrality,
respectively, and
\begin{eqnarray}
q_L= 0.8 {\rm ~GeV}, \, f_R \langle N_K \rangle = 3.0. 
\end{eqnarray}
for the PHENIX data with 0-20\% centrality.
In Figs.\ 1-4, we compare the experimental solid data points
\cite{Ada05,Put07,Wen08,Ada08} for the associated particle momentum
distribution with theoretical momentum kick model results shown as the
solid curves.  There is a general agreement between the theoretical
results and experimental data over an extended region of $p_t$,
$\Delta \phi$, and $\Delta \eta$, indicating the approximate validity
of the momentum kick model.  In these figures, we also show the
contributions from the ridge component as dashed curves.  The open
circles and the dash-dot curves give the experimental associated
particle data with the corresponding parametrized theoretical fit for
$pp$ collisions.
\begin{figure} [h]
\includegraphics[angle=0,scale=0.40]{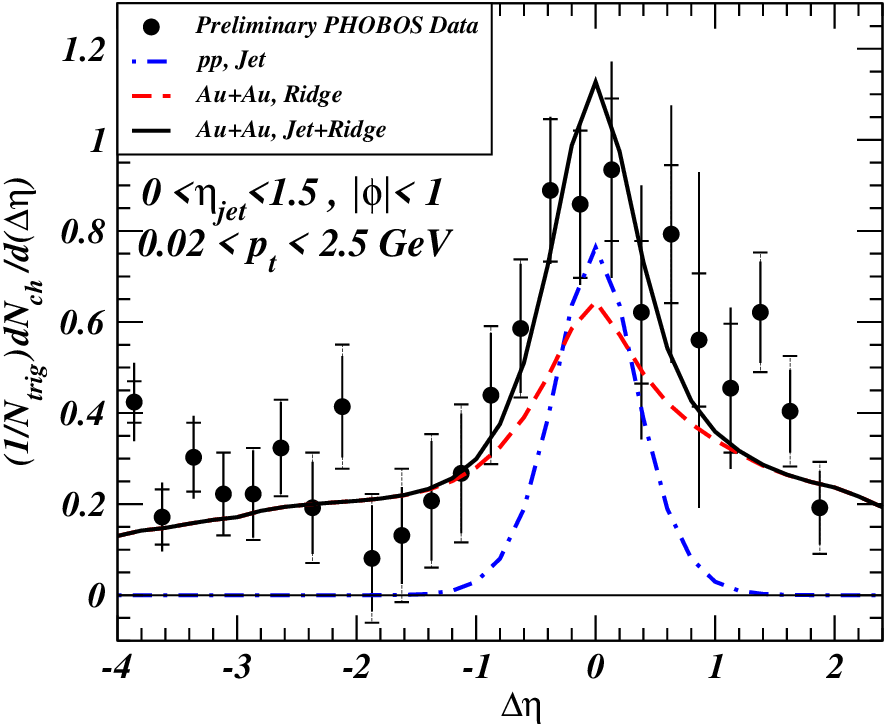}
\vspace*{0.0cm}
\caption{  The symbols represent PHOBOS experimental data
\cite{Wen08} and the curves theoretical results of $dN_{\rm
ch}/N_{\rm trig}d \Delta \eta$, for central AuAu collisions.
The dash-dot curve give the theoretical $pp$ jet yield.  }
\end{figure}
\begin{figure} [h]
\includegraphics[angle=0,scale=0.40]{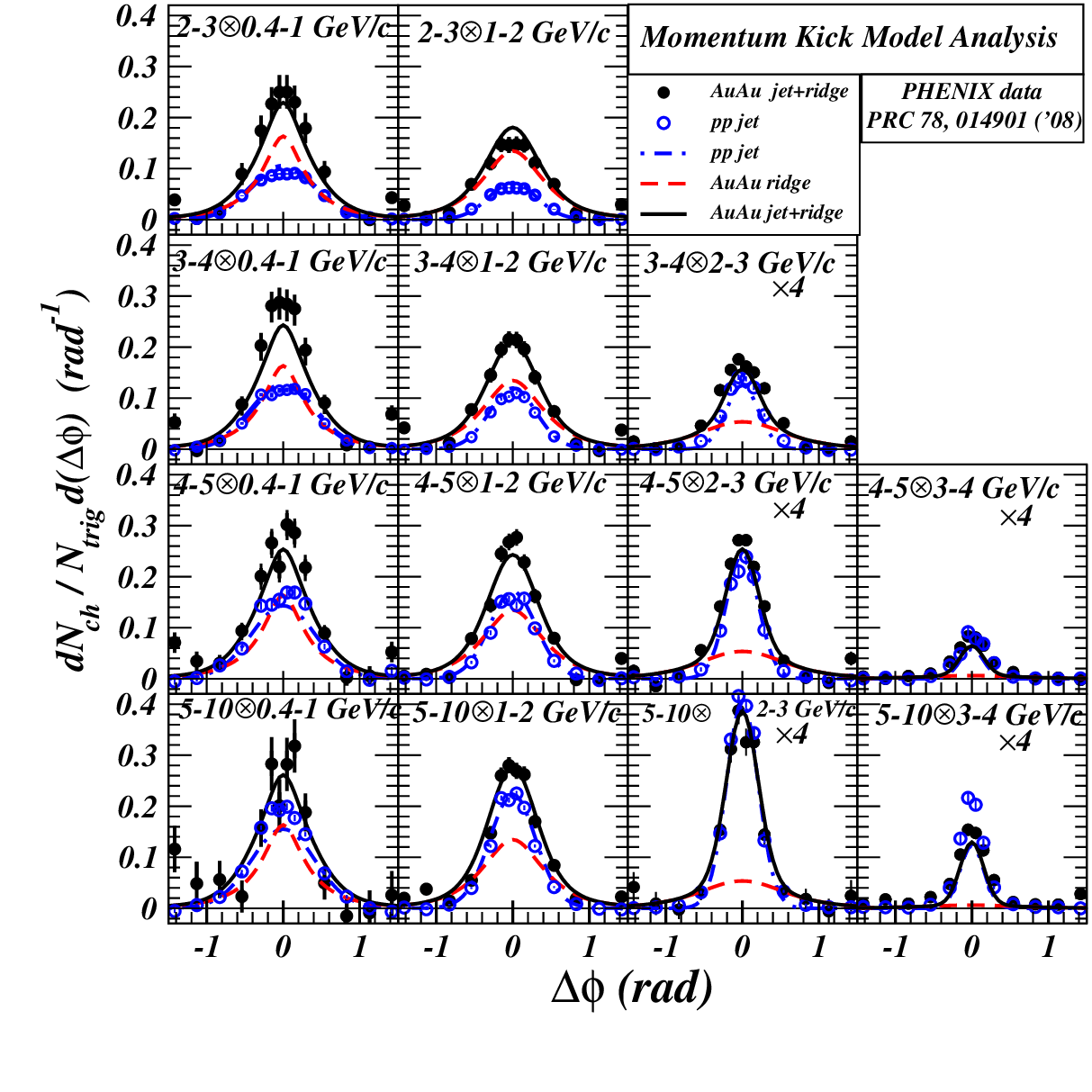}
\vspace*{-0.5cm}
\caption{The symbols represent PHENIX experimental data
\cite{Ada08} and the curves theoretical results of $dN_{\rm
ch}/N_{\rm trig}d\Delta \phi$, for $pp$ and central AuAu collisions.}
\end{figure}

\section{Information Extracted from the Momentum Kick Model}
With the help of the momentum kick model, the comparison with
experimental data furnished the following pieces of valuable
information:
\begin{enumerate}
\vspace*{-0.2cm}
\item
 The extracted early parton momentum distribution has a thermal-like
transverse momentum distribution and a rapidity plateau structure, as
represented by the parameters $a,T,$ and $m_d$ and shown in Fig. 5.
The rapidity plateau extends to large rapidities and gives rise to the
flat distribution of the associated particles as observed by the
PHOBOS Collaboration.  It is narrower than the $pp$ rapidity plateau
but wider than the Au-Au Gaussian structure \cite{Yan08,Won09a}, and
places the early partons to be at an intermediate stage of the
parton rapidity evolution. 
\vspace*{-0.2cm}
\item
The inverse slope $T$ of the early parton transverse momentum
distribution is intermediate between those for the $pp$ jet $T_{\rm
jet}$ and $T_{\rm bulk}$ for central Au-Au collisions \cite{Put07}.
This again indicates the intermediate nature of the early partons in
the evolution of the parton momentum distribution.
\vspace*{-0.2cm}
\item
The longitudinal momentum transfer by the jet onto the medium parton
in a collision is $q_L$=0.8-1.0 GeV.  Such a longitudinal momentum
transfer corresponds to a $|t|$ value of about 0.26 GeV$^2$, which
places the jet-parton collision within the realm of non-perturbative
parton-parton scattering, as in the non-perturbative scattering with
the exchange of a Pomeron.  The transverse correlation length $a$
extracted from the momentum kick model is compatible with those from
other non-perturbative treatment of parton-parton scattering as an
exchange of a Pomeron \cite{Won09a}.
\begin{figure} [h]
\includegraphics[angle=0,scale=0.40]{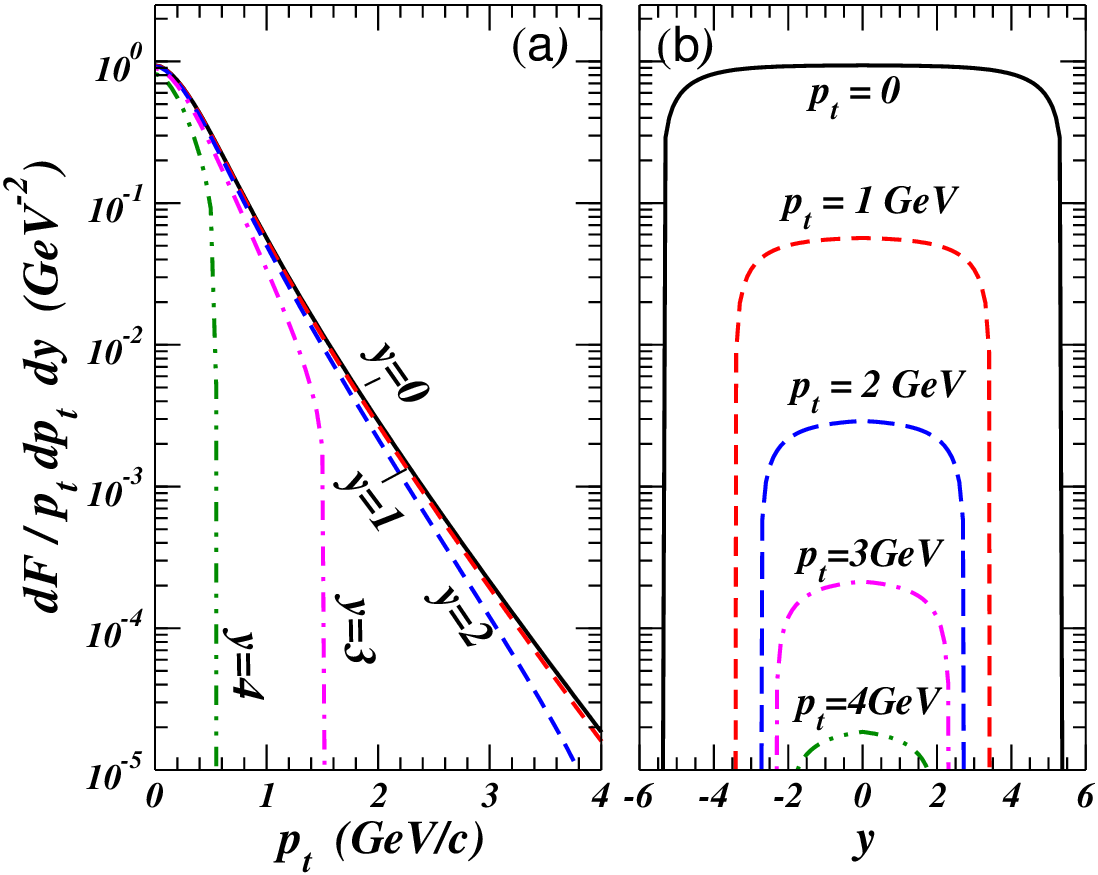}
\vspace*{0.0cm}
\caption{ Normalized initial parton momentum distribution $dF/dy p_t
dp_t$ extracted from experimental ridge data on the near-side ridge
data from the STAR, PHOBOS, and PHENIX Collaborations
\cite{Ada05,Put07,Wan07,Wen08,Ada08}. (a) $dF/dy p_t dp_t$ as a
function of $p_t$ for different $y$, and (b) $dF/dy p_t dp_t$ as a
function of $y$ for different $p_t$.  }
\end{figure}
\vspace*{-0.2cm}
\item
For the most central Au-Au collision, the average number $\langle
N_k\rangle$ of medium partons kicked by the jet on the near-side,
multiplied by the survival factor $f_R$ is $f_R \langle N_k\rangle =
3.0-3.8$.  The trend of the decrease of the number of kicked medium
partons and the degree of jet quenching as a function of centrality
are consistent with this simple picture of the momentum kick model
\cite{Won08a}.
\end{enumerate}

\section{ Conclusions and Discussions}

The characteristics of the associated particle experimental data
reveal that particles associated with a near-side jet arise from two
different components.  The ridge component can be interpreted as
arising from the medium partons which suffer an elastic scattering
from the jet passing through the dense matter.  The jet component can
be described as an attenuated distribution from a $pp$ collision.
Such a picture provides a good description of the experimental
associated particle data from the STAR, PHOBOS, and PHENIX
Collaborations.

The momentum kick model analysis allows us to extract a wealth of
valuable information on the states of the matter formed at the early
stage of the collision process.  We find that the momentum
distribution of the early partons at the moments of jet-parton
collisions has a rapidity plateau structure and a thermal-like
transverse momentum distribution.

Rapidity plateau structures appear in many multi-particle production
processes. Theoretically, a rapidity plateau is expected in QED2
fragmentation in a flux tube, which mimics particle production in QCD
as a $q$ and a $\bar q$ pull away from each other at high energies
\cite{Sch62,Cas74,Won91,Won94}.  Experimentally, a rapidity plateau
has been observed in high energy $e^+$$e^-$ annihilations, and $pp$
collisions \cite{Aih88,Yan08}.  Thus, the possibility of a rapidity
plateau for the ridge particles should not come as a surprise.
However, the fact that a parton of large absolute rapidity can occur
together with a jet of central rapidities, as revealed by the PHOBOS
data and the momentum kick model analysis, poses an interesting
conceptual question.  For those medium partons with large magnitude of
the longitudinal momentum to collide with the jet so as to become an
associated ridge particle in the PHOBOS experiment, the partons must
be present in the longitudinal neighborhood of the jet at the moments
of the jet-medium collision, at an early stage of the nucleus-nucleus
collision.

In many classical descriptions of particle production processes such
as the Lund model, a particle with a large absolute rapidity is
associated with a large separation from the longitudinal origin at a
later time.  Jet parton and medium parton collisions take place at an
early stage. Partons with large rapidities may not be produced for the
jet to collide.

It is important to point out that there are important quantum effects
\cite{Sch62,Won09b} that are beyond the realm of such classical
considerations.  The space-time dynamics of produced particles in
string fragmentation can be obtained by evaluating the Wigner function
of the produced particles when a strongly-coupled fermion separates
from an antifermion at high energies in QED2.  We find that produced
particles with different momenta in different regions of the rapidity
plateau are present at the moment when the overlapping
fermion-antifermion pair begin to separate \cite{Won09b}, in contrast
to the classical description of particle production.  Thus, the
momentum kick model is consistent with other knowledge of the
interaction processes and supports the picture of color flux tube
production of the medium partons at an early stages, as in Schwinger's
non-perturbative QED2 string fragmentation.

\label{last}
\end{document}